\documentclass[AMA,STIX1COL]{WileyNJD-v2}
\usepackage{algorithm}
\usepackage{algpseudocode}
\usepackage{graphicx,caption}
\usepackage{amsmath}
\usepackage{tikz}
\usepackage{calc}
\usepackage{amssymb}
\usepackage{pifont}
\articletype{Article Type}%

\begin{document}

\title{Smart Contract Assisted Blockchain based PKI System}

\author[1]{Amrutanshu Panigrahi*}

\author[2]{Ajit Kumar Nayak}

\author[3]{Rourab Paul}

\authormark{Amrutanshu Panigrahi \textsc{et al}}

\address[1]{\orgdiv{Department of CSE}, \orgname{Siksha O Anusandhan (Deemed to be University)}, \orgaddress{\state{Odisha}, \country{India}}}

\address[2]{\orgdiv{Department of CS\& IT}, \orgname{Siksha O Anusandhan (Deemed to be University)}, \orgaddress{\state{Odisha}, \country{India}}}

\address[3]{\orgdiv{Department of CSE}, \orgname{Siksha O Anusandhan (Deemed to be University)}, \orgaddress{\state{Odisha}, \country{India}}}

\corres{*Amrutanshu Panigrahi \email{amrutansup89@gmail.com}}

\presentaddress{Siksha O Anusandhan (Deemed to be University)}

\abstract[Summary]{Public Key Infrastructure (PKI) is a reliable solution for Internet communication. PKI finds applications in secure email, virtual private network (VPN), e-commerce, e-governance, etc. It provides a secure mechanism to authenticate users and communications. The conventional PKI system is centralized, which exposes the infrastructure to many security issues. The digital certificate generation and validation processes in PKI suffer from high latency and inadequate authentication processes. Moreover, it needs enormous time and effort to mitigate the malfeasance of the Certificate Authority (CA). The complexity of employing the traditional key and certificate management increases by enforcing the centralized $CA$, which can compromise the transaction security. To overcome the aforementioned issues of PKI, three different solutions have been reported in the literature: Log based PKI (LBPKI), Web of Trust (WoT), and blockchain based PKI. The blockchain based PKI achieves more attention as it is the combination of LBPKI and WoT, which serves distributed trust, log of transactions, and constant sized data to verify the identity of users. Motivated by these facts, this article reports a blockchain-based PKI system which has a lighter smart contract and less storage capacity and is also suitable for lightweight applications. The lighter smart contract in our infrastructure uses a $threshold~value$, which validates the limit of one participating node for becoming the $CA$ of any transaction inside the network. This approach can prevent distributed denial of service (DDoS) attacks. This smart contract also checks the signer node address. The proposed smart contract can prevent seven cyber attacks, such as Denial of Service (DoS), Man in the Middle Attack (MITM), Distributed Denial of Service (DDoS), 51\%, Injection attacks, Routing Attack, and Eclipse attack. The Delegated Proof of Stake (DPoS) consensus algorithm used in this model reduces the number of validators for each transaction which makes it suitable for lightweight applications. The timing complexity of key/certificate validation and signature/certificate revocation processes do not depend on the number of transactions. The comparisons of various timing parameters with existing solutions show that the proposed PKI is competitively better.}

\keywords{PKI, WoT, LBPKI, Smart Contract, Blockchain based PKI}
\maketitle

%\footnotetext{\textbf{Abbreviations:} ANA, anti-nuclear antibodies; APC, antigen-presenting cells; IRF, interferon regulatory factor}

\section{Introduction}\label{sec1}

PKI is the primary building block of client-server communication over the internet. PKI defines a set of rules and protocols for the crypto algorithms: encryption, decryption, digital signature, and digital certificate verification process, which are used in secure communication. For server identity authentication, traditional PKI uses a digital certificate which is issued by a trusted third party named as Certificate Authority ($CA$). This certificate is a data package to identify the identity of the server. The digital certificate is associated with the public key, and it is protected by asymmetric key cryptography. The $CA$ has three primary responsibilities (i) issuing, (ii) revoking (iii) distributing digital certificates. Therefore, it is the most crucial component of PKI. The digital certificate standard, ITU-T X.509\cite{li2020survey} coheres to the public key with the DNS record. The X.509 standard certificate provides a verification method for the private and public keys used for the communication. $CA$ is the only component in  PKI to validate a transaction. Traditional PKI system adopts a trusted third party for issuing the digital certificate for every transaction or communication over the internet. There are various third-party $CA$s reported in the literature, such as Comodo, IdenTrust, DigiCert, Certum, Entrust, etc\cite{wen2021attacks}. The degree of the successful transaction between the client and server depends upon the correctness of the certificate issued by $CA$. The communications in the aforementioned PKIs rely on the third-party centralized $CA$s. If the $CA$s used in Comodo, IdenTrust, DigiCert, Certum, Entrust, etc., become malicious, then the entire communication will be compromised, and it leads to single point failure \cite{singh2021blockchain}. Comodo is the first $CA$ which have suffered from cyberattacks. In 2011 it had issued nine fraud digital certificates to various domains. In the same year DigiNotar has issued around 600 fraud certificates to various organizations \cite{lu2019blockchain}.\par 

Despite single point failure \cite{almasoud2020smart}, the conventional PKI system has several other drawbacks. The conventional PKI does not have any feature to detect compromised $CA$.

 Moreover, the complexity of key generation and key validation processes reduces the performance of the conventional PKI. Considering these threats, servers which are not able to secure their own identities satisfactorily cannot ensure that their communications are not compromised by a deceitful certificate which may cause Man in the Middle attack (MITM) \cite{gourisetti2019evaluation}. 
\par The malicious certificate issued by a compromised $CA$ can cause severe damage to the transactions of conventional PKI. A malevolent $CA$ like in DigiNotar loses all of its trustworthiness, and it creates a rogue certificate, which makes the entire network at risk \cite{singh2021blockchain}.
Therefore, the aforementioned statements brief four major concerns of conventional PKI:
\begin{itemize}

\item The trust of existing PKI is centralized to  Certificate Authority (CA) which can cause single point failure.
\item The communications governed by PKI rely on the third-party centralized CAs. The literature has reported many incidents of malicious CAs. 
\item There are no ways to detect malicious CA.
\item The complexity of key generation and key validation processes reduces the performance of the conventional PKI.

\end{itemize}
 Pretty Good Privacy (PGP)\cite{xu2020segment} is one of the cryptographic solutions against the issues stated above. Unlike traditional $CA$, PGP gives the opportunity to the participating node to verify the digital certificates of other participating nodes by including their corresponding signature. This attribute creates a trust model where every participating node becomes the verifier for the other. As stated above, the issues of conventional PKI systems are properly addressed by 3 different approaches, such as $Web~of~Trust$, $Log~based$, and $Blockchain~based$ \cite{xiong2020data}. \par

Web of Trust (WoT) is the first approach which addresses the centralization issue of conventional PKI. WoT allows the network participants to choose their own trustworthy certificate provider for transactions. This feature decentralizes the whole infrastructure. The crucial drawback of the WoT is the overhead of the new joinee. The selection process of $CA$ in WoT network is very complicated, which makes it inappropriate for conventional applications. At each successful transaction, the $CA$ increases its trust counter value. Thereafter, for the next transaction, the node chooses a validator which has the highest counter value. The counter value of a new joinee in WoT network is zero. Therefore, the new joinee will never be selected as a validator of any transaction. This issue makes WoT unrealistic for PKI applications \cite{ali2021comparative}.
\par Public log used in Log Based PKI is one of the solutions which can monitor activities of the $CA$. The log server will be visible to the entire network. Any illegitimate digital certificate can be identified by this network, and the corresponding $CA$ will be suspended due to its malicious activity \cite{ali2021comparative}. The public log server used in Log Based PKI is always prone to single point failure issue, which is the main disadvantage of this infrastructure \cite{paik2019analysis}. The literature also provides many blockchain based PKIs, which are discussed in Sec. \ref{sec:rel} and Sec. \ref{sec:ps}.

%To overcome the above limitations, the smart contract-based PKI has been introduced. It simply dissociates the storage from the validation process where one node need not to store the entire blockchain copy for validating a transaction \cite{kfoury2020blockchain}.

%This paper concludes two primary issues of existing PKI systems. 
%\begin{itemize}
%\item Conventional PKI system is pruned to the single point failure due to which the entire network can be deactivated due to the CA
%\item Inappropiriate method to detect the malicious CA and rouge certficate in case of the conventional system.
%\end{itemize}
\subsection{Related Work}
\label{sec:rel}
In the current section, several PKI solutions are discussed. The discussion includes a PKIs without blockchain technology in Sec. \ref{without} and  blockchain-based PKI solutions in Sec. \ref{with}.
\subsubsection{PKI without Blockchain\cite{CT}\cite{kim2013accountable}\cite{basin2014arpki}\cite{POLI}\cite{dumas2017localpki}\cite{vigil2012notary}}\label{without}
This section discusses about existing PKIs frameworks which have not used blockchain.This type of PKI is further categorized into two groups: log based PKI (LBPKI) ( \ref{LBPKI1}) and WoT based PKI (\ref{WoTPKI2}).
\begin{enumerate}[A]
\item \textbf {LBPKI:}\label{LBPKI1}
Certificate Transparency (CT) in articles \cite{CT} maintains a public log of all issued certificates which strives to alleviate the problem of incorrectly issued certificates. The public logs are auditable. Therefore, it is easier for any nodes to check different activities like new certificates generation and certificate deletion. The public logs do not eliminate the risk of certificate misuse. It does not guarantee that the user is able to notice certificate misuse when it occurs.\par
Proposed Accountable Key Infrastructure (AKI) \cite{kim2013accountable} is used to defend domains and clients from flaws induced by single points of failure. The check and balance method in AKI distributes the trust properly among multiple parties including $CA$s and domains. Even if the domain key is lost or breached, the AKI executes routine certification processes effectively and gracefully. 
It was presented as a solution for a public-key validation infrastructure. It selects a set of trusted nodes for validating the entire transactions in the network which decreases the dependency on any one node.
Attack Resilient Public-Key Infrastructure (ARPKI)\cite{basin2014arpki} makes all of the certificated-related computations such as (i)certificate issue, (ii)update, (iii)revocation, and (iv)validation processes transparent. ARPKI starts working with 2 different parts. The first part contains two different CAs and the second part contains one Integrated Log Server (ILS) for performing any operations. It ensures that the security will be preserved, even if the $n-1$ nodes are compromised out of all $n$ number nodes.\par
Policert\cite{POLI} is a broad log-based and domain-oriented architecture which uses a more secure authentication process for securing the domain's public keys and an extensive certificate management method for validating the transaction.
\item \textbf{WoT based PKI :}\label{WoTPKI2}
LOCALPKI \cite{dumas2017localpki} was developed for the Internet of Things applications. In this PKI a local authority binds the public key with the user identity and the certificate is issued by a third-party node or local authority. A third-party entity is used in LOCALPKI to record this binding information and to provide registration updates.\par
The Notary-based PKI (NBPKI) \cite{vigil2012notary} approach creates a group of trustworthy individuals known as Notarial Authorities (NA). The NA confirms the reliability of a certificate for validating a certain signature at a specified time.  The end users depend on NA's public keys and self-signed certificates for producing and validating signatures. The working principle of NBPKI relies on three different components (i)end-user, (ii)Registration Authority (RA), and the Notarial Authority (NA). The end-user needs to register with RA for signing their transactions. The RA verifies the end-user identity and informs the associated NA. The NA decides the status of the trustworthiness of the end-user based on the information provided by the RA.
\end{enumerate}
\subsubsection{Bloockchain based PKI \cite{wang2018privacy}\cite{wang2018blockcam}\cite{hammi2018bctrust}\cite{yakubov2020blockpgp}\cite{axon2016pb}\cite{ahmed2018turning}\cite{chen2018certchain}\cite{kubilay2019certledger}\cite{toorani2021decentralized}\cite{matsumoto2016ikp} \cite{bunz2020flyclient}\cite{exosite2019blockquick}}\label{with}
This paper primarily addresses 8 attributes to compare different PKI system such as feature, type of blockchain network, blockchain platform, certificate, trust model, off-chain storage, on-chain and time complexity.
 Table \ref{tab:comp} shows the detailed study of different blockchain based PKI systems.
\begin{itemize}
\item \textbf{Key Feature:} It shows the basic characteristic such as smart contract, CA, public ledge, etc. The blockchain based PKI is developed based on these key features.
\item \textbf{Blockchain type:} The adopted blockchain network can be either of permissioned or permissionless blockchain. In a permissioned network, the new node can only join when it gets permission from every participating node present in the network whereas, in the permissionless network, new nodes do not require permissions from other nodes exist in the network. Instead of that, it takes permission either from one trusted node or from anyone randomly chosen node.
\item \textbf{Blockchain Platform:} It shows the platform on which the PKI is implemented. The platform can be on the shelf platform such as Ethereum or a self-developed custom platform. The shelf platforms are publicly available and it needs to be downloaded from a trusted source and configured as per the requirement.
\item\textbf{Certificate:} It shows the type of certificate used during the PKI development. It can be a X.509 standard or a custom one.
\item\textbf{Trust Model:} It represents the mechanism for selecting the $CA$ for validating a transaction. One node can choose a trustworthy node or a random node who solves the NONCE first.
\item\textbf{Consensus Model:} It shows the adopted consensus model during the PKI development.
\item\textbf{Storage:} The blockchain data can be stored in two forms such as the entire copy of the data will be stored, or the hash function of the block will be stored. There are two categories present for blockchain data storage named as on-chain storage and off-chain storage. On-chain storage allows the node to store the data directly on the blockchain network. Whereas the off-chain storage allows storing the data in a public ledger that is accessible by all other nodes or in a private storage from which that particular node can access it.
\item\textbf{Time Complexity:} This shows the algorithmic computational complexity in terms of time. It has been taken in big O format as for every PKI all of the defined methods needed to be executed for a successful transaction. So the worst time complexity has been considered for different available blockchain PKI.
\end{itemize}
\begin{table}[htb]
\centering
\caption{Comparative study of existing blockchain based PKI systems based on the defined features}
\label{tab:comp}
%\scriptsize
\resizebox{16cm}{!}{  

\begin{tabular}{|p{1.9cm}|p{1.9cm}|p{1.7cm}|p{1.2cm}|p{1.4cm}|p{1.6cm}|p{1.4cm}|p{1.1cm}|p{0.8cm}|p{1.5cm}|}
\hline

PKI      & Key Feature                                 & Blockchain Type & Block-chain Platform & Certificate & Trust Model  & Consensus Model & Off-chain Storage & On-chain Storage & Time Complexity  \\ 
\hline
PA-PKI \cite{wang2018privacy} & Identification and verification of CA	 & Permissioned    & Ethereum               & Custom    & WoT             &	PBFT                 &	Private            &	Hash	             & $O(n)$	                 \\
\hline
Block-CAM	\cite{wang2018blockcam} & $CA$ in cross domain verification	& Consortium and Permissioned  &  Ethereum & X.509 v3 & Hierar  chical	 & NA & Public Data & Hash + Data & $O(n^{3})$  \\
\hline
BC-TRUST	\cite{hammi2018bctrust} & Authentication	& Permission Less &	Ethereum &	Custom	& WoT	& NA	& Public Data &	Hash + Data	& --	 \\
\hline
BLOCK-PGP \cite{yakubov2020blockpgp} &	Access Control of Certificate Revocation &	Permissioned	& Ethereum &	X.509 &	WoT &	PoW	& Public Data &	NA	& --  \\
\hline
PB-PKI \cite{axon2016pb} &	Public Ledger &	Permission Less	& Custom	& Custom	& WoT &	NA &	Private Data &	Hash	& O(n)  \\
\hline
TTA-SC \cite{ahmed2018turning} &	Automating the process of identifying the misconfigured $CA$ &	Permission Less &	Ethereum &	X.509	& Hierarchical	& NA &	Public Data &	Hash + Data &	$O(n)$ \\
\hline
CERT-CHAIN \cite{chen2018certchain}	& $CA$ Trustworthy by using Dual Counting Bloom Filter (DCBF) &	Permission Less	 & Custom &	X.509	& Hierar-chical &	Dependa-bility rank based &	Public Data &	Hash	& $O (n^{2}log(n))$  \\
\hline
CERT-LEDGER \cite{kubilay2019certledger} &	Certificate Transparency &	Permission Less &	Ethereum &	X.509 &	Hierarchical &	PBFT &	Public
Data	& Hash	& $O(log(n))$  \\
\hline
DB-PKI \cite{toorani2021decentralized} &	$CA$ & 	Permission Less &	Custom &	Custom &	WoT &	PBFT	 &Public Data &	Hash &	$O(n^2)$  \\
\hline
IKP	\cite{matsumoto2016ikp} & $CA$ trustworthy &	Permission Less	& Ethereum &	X.509 &	Hierarchical &	NA	& Public &	Hash +Data &	$O(nlog(n))$  \\
\hline
FLY-CLIENT \cite{bunz2020flyclient} & Transaction Verification for light client &	Permissioned	& Ethereum &	Custom &	Hierarchical	& PoS	& Public &	NA	& $O(logn)$	 \\
\hline
BLOCK-QUICK \cite{exosite2019blockquick}	&Transaction Verification for light client &	Permission Less	& Ethereum	& Custom	& WoT &	PoPoW	& Public &	NA &	$O(n)$	 \\
\hline
\end{tabular}

}
\end{table}

\subsection{Problem Statement and Motivation}
\label{sec:ps}
The trust of the traditional PKI systems completely depends on third-party $CA$s. The $CA$ checks the bindings between public keys and entities and then provides digital certificates to those entities. A digital certificate assures that a $CA$ confirms the binding process \cite{zhu2020guided}. There are a very limited number of $CA$s which are trusted by the modern browser and OS manufacturers. Therefore, this CA-based PKI architecture is considered a centralized infrastructure. The present CA-based PKI architecture, such as CT \cite{CT}, AKI \cite{kim2013accountable}, and ARPKI \cite{basin2014arpki} have adopted many methods to reduce the dependence on the confidence of $CA$.
 The primary concern in adopting those PKIs is to avoid the centralization issue of the infrastructure. \par
%Blockchain may be thought of as a distributed database whose content is agreed upon by network users via the use of a consensus mechanism. It is based on a peer-to-peer network in which members use the gossip protocol to disseminate the current state of the blockchain. For the reasons stated above, blockchain-based PKIs are a trendy topic of research, with several proposed blockchain-based PKIs such as PA-PKI \cite{wang2018privacy}, Block CAM \cite{wang2018blockcam}, PB-PKI  \cite{axon2016pb}, etc. Though many blockchain-based solutions are there to avoid the issues of the traditional PKI system, most of the existing blockchain-based PKI system does not provide a fair chance to all present nodes to become the CA. Instead, the main focus is on identifying the CA misbehavior. Exiting blockchain-based PKI solutions avoids the security aspect of the network i.e., how to avoid various kinds of attacks to the network.

\par Blockchain Based PKIs such as PA-PKI \cite{wang2018privacy}, Block CAM \cite{wang2018blockcam}, PB-PKI  \cite{axon2016pb} etc. provide an emerging alternative for conventional PKI system which adopts different features of Log based and WoT approaches. Blockchain based PKI provides an environment for decentralized authentication and validation of transactions in the network \cite{omar2021implementing}. The adoption of different $CA$s for different transactions in Blockchain based decentralized PKIs eliminates many issues caused by legacy PKIs. The use of different $CA$s for different transactions increases the fault tolerance capacity of the network and one malicious $CA$ can not sabotage the entire chain.
The distributed log in blockchain-based PKI provides a certificate transparency feature which is similar to the certificate transparency (CT) characteristic provided by Google which helps to improve the security of PKIs. The CT allows logging and observing the scope of digital certificates. The examples of blockchain based PKI systems are Namecoin and Emercoin \cite{karaarslan2018blockchain}. The Namecoin and Emercoin need enormous storage for the entire blockchain information for validation purposes and they also need to store the entire blockchain copy at the user end. These storage issues have made these blockchain based PKI impractical for real life applications. The smart contract-based PKI simply dissociates the storage from the validation process where one node does need not to store the entire blockchain copy for validating a transaction \cite{kfoury2020blockchain}. The major lacunas of existing blockchain based PKIs are :
\begin{itemize}
\item All the participants in existing blockchain based PKI do not get a fair chance to become $CA$. 
\item This complexity of the consensus algorithm in blockchain based PKI makes it inefficient specially for the lightweight application.
\item Most of the blockchain based PKIs have concentrated on Denial of Service (DoS) and Man in the Middle Attack (MITM). They have not addressed Distributed Denial of Service (DDoS), 51\% attack, Injection attacks, Routing Attack and Eclipse attack.
\end{itemize}

\subsection{Contribution}
The proposed smart contract-based PKI addresses the challenges of existing blockchain based PKIs. The contribution of the research article is summarized as follows.\par
\begin{itemize}
\item The proposed smart contract of Blockchain based PKI can prevent DoS, DDoS, MITM, 51\%, Injection, Routing, and Eclipse attacks. The proposed smart checks the validity of the signer node address and it also imposes a threshold value for becoming $CA$ which gives a fair chance to all the participants to become $CA$.
\item This paper adopted Delegated Proof of Stake (DPoS) consensus algorithm which reduces the number of validators of each transaction. Therefore it reduces the timing complexity which makes it suitable for lightweight applications.
\item The proposed PKI system is evaluated based on the two matrices. (i)lapse time of key generation and key validation process and (ii) gas cost of the transaction. The result shows the time complexity of the proposed blockchain based PKI system is efficient compared to existing literature.
\end{itemize}

\subsection{Structure}
The rest of the paper is structured in the following manner. Section \ref{prilim} focuses on the background study for the current work. The proposed blockchain based PKI system based on the smart contract is presented in Sec. \ref{proposed}. The working principle of the proposed PKI system is elaborated in Sec. \ref{work}. The proposed model is evaluated based on the gas cost and latency for key generation and validation in Sec. \ref{implement}. Finally, Sec. \ref{conclusion} represents the conclusion and the future scope of the research work.

\section{Preliminaries}\label{prilim}
Various preliminary elements like the notion of blockchain, smart contracts, and the fundamental concept of PKI are reported in this section.
\subsection{Blockchain}
In 2008, a whitepaper \cite{nakamoto2008bitcoin} was released under the pseudonym of "Satoshi Nakamoto" who is considered as the pioneer of the idea of blockchain. Along with the popularity of Bitcoin, blockchain has sparked a lot of interest in academic research and practical applications. The information stored in the immutable ledger of the blockchain can be read by any node inside the network. It combines technologies such as consensus algorithm, smart contract, peer-to-peer (P2P), and encryption to create a new distributed computing paradigm \cite{zheng2017overview}.\par
Decentralization, immutability, and transparency features of blockchain overcome many problems like high cost, poor efficiency, and single point of failure \cite{kumar2020secure}. Blockchain 1.0, reported by Bitcoin, was created only for the decentralization of cryptocurrency. Since then, many flaws have been addressed in the literature. Non-Turing completeness is an issue, which limits the blockchain's use in many other fields. Blockchain 2.0 is mostly represented by Ethereum which introduced a smart contract feature. Smart contracts are based on distributed architecture and consensus methods, which allow transactions among users without mutual trust. As a result, smart contracts based blockchain has a lot of potential \cite {bhushan2021untangling} in existing applications.
\subsection{Smart Contract}
Externally Owned Addresses (EOA) and smart contract accounts are the two forms of Ethereum accounts. Users own EOA, which includes private key-public key pairs. A smart contract is a program that does not have any key pair and it checks all contract criteria \cite{wang2019blockchain} required for that specific application. It includes one or more trigger conditions, such as a specified time or occurrence, and the related reactions, such as a specific transaction or activity \cite{zheng2020overview}. Once it is signed by all parties, the contract is linked to the blockchain data. These contracts are propagated across the P2P network, confirmed by all nodes, and finally deposited in a particular block of the blockchain. Users who know the address of the contract, interface, and other certain details call the smart contract while transactions are started \cite{bartoletti2017empirical}. The miner runs the contract code in the local sandbox environment after getting a call message about the transaction. If the given contract criteria are matched, then all the defined operations for that transaction will be executed \cite{wang2021security}. Once the transaction is validated, it is authenticated and inserted into a new block using a consensus procedure. The blockchain also stores the transaction and the modified state along with the status of the participants\cite{patsonakis2020implementing}.
\subsection{Ethereum}
Ethereum is an open-source platform where the smart contract is the key functionality. It provides a virtual environment where multiple live nodes are deployed to create a blockchain network. Smart contracts are written in Turing complete language known as Solidity which is executed in the Ethereum virtual machine \cite{zhang2020btcas}. The smart contract code is publicly available to all participating nodes present in the blockchain network. In the current research work, multiple nodes are deployed with some initial cost and gas using GETH. For every transaction, the node needs to share some gas ($G$) and each gas has some price ($P$). So, the total cost ($C$) in terms of ether (ETH) can be expressed as equation\ref{eq1}. 
\begin{equation}\label{eq1}
C=G \times P 
\end{equation}
\subsection{PKI}
Certificate authority (CA), registration authority (RA), certificate revocation list (CRL), central directory (CD), lightweight directory access protocol (LDAP), and online certificate status protocol (OCSP) are all components of a typical PKI system \cite{adja2021blockchain}.
\begin{itemize}
\item\textbf{CA:} In the PKI system, $CA$ is a trusted central party. It is in charge of disseminating public keys and issuing certificates using the CA's private key to validate the transaction. The certificates are kept in a repository for future searches and verification. Normally, $CA$ trust is organized in a chain/tree structure with multiple levels. Root $CA$ (RCA), intermediate $CA$ (ICA), and signing $CA$ (SCA) are all included from top to bottom. The RCA is in dominating position, which may be self-certified. Certificates for other sub-CAs may be issued by root CAs. 
\item\textbf{RA:} The registration of users is the responsibility of RA. When a user requests a certificate, RA must verify the information that the user has submitted. In compact PKI systems, the RA and $CA$ may be the same entity, or a RA might be an ICA or SCA. In large PKI systems, the $CA$ may also select additional trustworthy parties as RAs.
\item\textbf{CRL:} All revoked certificates signed by the $CA$ are stored in the CRL. In addition, the user must additionally search the  CRL to determine the status of a certificate for the validation process.
\item\textbf{CD:} It is a trustworthy server that is used to store the issued certificate in contrast to the corresponding key.
\item\textbf{LDAP:} It is a protocol that allows users to quickly access a certificate storage repository which may be CD or CRL.
\item\textbf{OCSP:} It is a protocol that becomes active when a user request to check the status of a certificate for validation from CRL.
\end{itemize}

%\begin{eqnarray}
%s(nT_{s}) &= &s(t)\times \sum\limits_{n=0}^{N-1} \delta (t-nT_{s}) \xleftrightarrow{\mathrm{DFT}}  S \left(\frac{m}{NT_{s}}\right) \nonumber\\
%&= &\frac{1}{N} \sum\limits_{n=0}^{N-1} \sum\limits_{k=-N/2}^{N/2-1} s_{k} e^{\mathrm{j}2\pi k\Delta fnT_{s}} e^{-j\frac{2\pi}{N}mn}
%\end{eqnarray}
\section{Proposed Work} \label{proposed}
The proposed smart contract based PKI system is implemented in the open-source Ethereum platform known as the Go Ethereum or GETH. The main building blocks of the proposed PKI system are smart contract and Ethereum. Ethereum is used as the platform where the smart contract is the core part of the work. 
\subsection{Model Description}The proposed PKI system contains three basic modules such as Participant, Smart Contract, Signature, and Revocation. The participant module contains the method to add the attributes of a participating node when it is new to the network. The signature module enables the nodes to sign and validate the keypair. The revocation module allows the node to revoke its own signature so that the corresponding node can resign another transaction. 
\begin{enumerate}[(A)]
\item\textbf{New Participant:}
The input of this module is the status of the node. If the node is found as a new node of the network, then the 3 attributes: $PID$, $ETH~address$ and $Keypair$ will be set to the status of the new node to participate in the transactions of the network. If a node already exists in the network, the participant module invokes the aforementioned attributes to participate in the transaction. The pseudo-code for this module is presented in algorithm \ref{algo:part}. The attributes  of the participant module are stated below:
\begin{itemize}
\item\textit{\textbf{PID:}}It is a unique random number that can be used to identify a particular node in the network.
\item\textit{\textbf{ETH address:}} It is an address provided by the Ethereum blockchain environment which is required during transactions.
\item\textit{\textbf{Keypair:}} The private and public key pairs will be generated and assigned to a particular node.
\end{itemize}
As the current research considers a lighter smart contact, only the PID of that corresponding node is stored after deployment.
\item\textbf{Smart Contract:} The inputs to this module are the $PID$, $RID$ and $ETH adress$. The $PID$ and the $ETH address$ of the chosen signer node are compared with the stored $PID$ and $ETH address$. If both of the addresses are matched then the $RID$ of the signer node will be compared with the defined threshold for that node. The transaction will be allowed only after the successful execution of the above said conditions. The detail pseudocode is reflected in algorithm \ref{algo:smart}.
\item\textbf{Signature Validation:}This module allows the nodes to sign the transactions of the other nodes. When the node is elected as the signer node, this method will be called with two attributes such as the PID and Expiry. The steps are shown in algorithm \ref{algo:sign}.
\begin{itemize}
\item\textit{\textbf{PID:}}It is the unique number assigned by the Participant method which provides the unique identity.
\item\textit{\textbf{Expiry:}} After the validation process the node needs to increase the predefined counter by one to ensure that all of the participant nodes present in the network will get an equal chance to become the transaction lead. This counter value is the maximum number for which one node can be elected as the transaction lead. In the current research work, it is defined in the smart contract to avoid the DDoS attack.
\end{itemize}
\item\textbf{Revocation:} It is called by the leader node after every transaction. It contains the counter described in the signature module. The node increases the counter by one after every successful transaction. If the counter exceeds the maximum limit defined in the light version of the smart contract, the election process is rejected and the process is reinitiated. Revoke ID or $RID$ and Signer ID are two attributes present in this module. The pseudo-code for this module is represented in algorithm \ref{algo:revo}
\begin{itemize}
\item\textit{\textbf{RID:}}It is a counter which is increased by the leader node after the successful completion of the transaction.
\item\textit{\textbf{Signer ID:}} It is the id of the node which is going to validate the transaction. 
\end{itemize}
\end{enumerate}

For certificate revocation and key pair updating process, the participating node needs to broadcast its public key inside the network in the form of a transaction. All the nodes in the network store the data to maintain the distributed trust, therefore, the content of the mined block can only be changed when the maximum number of the participating node agree on the changed value.

\begin{algorithm}
\caption{New Participant}
\begin{algorithmic}
\State  BEGIN TRANSACTION
\State REQUIRE: \textbf {Set of Nodes N=[$N_{1}$,$N_{1}$,$N_{1}$,...........,$N_{n}$]}
\State PROC \textit{\textbf{PARTICIPANT}}()
\State get $N_{i}$.status
\If {($N_{i}$.status==FALSE)}
\State set PID
\State set (PR\textsubscript{key}, PB\textsubscript{key})
\State set PID.getETHAddress()
\State set PID.Limit
\Else
\State Node $N_{i}$ is present in the Ethereum private network
\EndIf
\State run CONTRACT()
\end{algorithmic}
\label{algo:part}

\end{algorithm}

\begin{algorithm}
\caption{Smart Contract}\label{algo:smart}
\begin{algorithmic}
\State {TRANSACTION PROCESSED}
\State REQUIRE: \textbf {$N_{i}$.RID,$N_{i}$.PID,$N_{i}$.ETHAddress}
%\vspace{-1em}
\State get Signer.PID
\State get Signer.RID
\State get Signer.ETHAddress
\If {(Signer.PID==$N_{i}$.PID $and$ (Signer.ETHAddress==$N_{i}$.ETHAddress)}
\If {($N_{i}$.RID $\leq$ $N_{i}$.limit)}
\State PROC SIGNATURE
\Else
\State Maximum Trial is over for the elected signernode. Please select another node
\EndIf
\EndIf

\end{algorithmic}

\end{algorithm}

\begin{algorithm}

\caption{Signature Validation}
\begin{algorithmic}

\State {TRANSACTION PROCESSED}
\State REQUIRE: \textbf {$N_{i}$.PID, $N_{i}$.ETHAdress}
\State PROC SIGNATURE ()
\State get Signer.PID
\State get Singer.ETHAdress

\State validate(TRANSCATION)
\State PROC REVOKE()
\end{algorithmic}
\label{algo:sign}
\end{algorithm}

\begin{algorithm}

\caption{Signature Revocation}
\begin{algorithmic}
\State {TRANSACTION PROCESSED}
\State REQUIRE: \textbf {$N_{i}$.RID}
\If {(TRANSACTION==TRUE)}
\State RID ++
\Else
\State  Transaction is rejected
\EndIf

\end{algorithmic}
\label{algo:revo}

\end{algorithm}

\subsection{Block structure}
\begin{figure}[h!]
\centering
\includegraphics[scale=0.4]{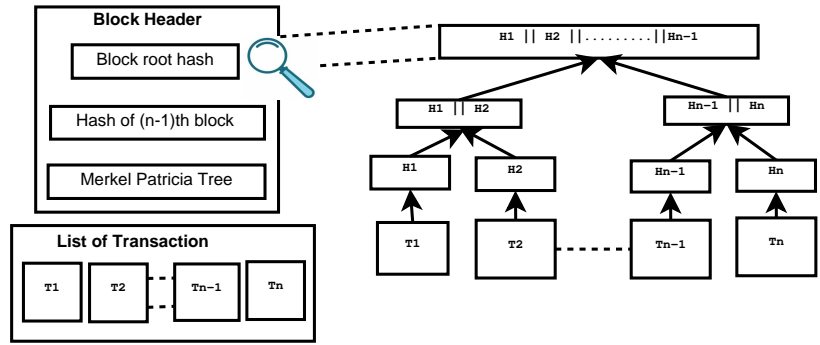}
\caption{Block structure of the proposed PKI} \label{fig:str}
\end{figure}

[H]
Each block has 2 components: block header and list of transactions.
Block header has 3 fields: (i) Block root hash, (ii) Hash of the previous transaction, and (iii) Markel Patricia Tree (MPT).
Figure \ref{fig:str} represents the structure of block where $n$ is the number of transactions. Here $T_1$ to $T_{n-1}$ are previous validated transactions and $T_n$ denotes the current transaction. $H_{i}$ denotes the hash value of $T_i$ where $i$ varies from $1$ to $n$. The number of transactions stored in a single block may vary with different blockchain platforms. The size of blocks on certain blockchains, such as Bitcoin, is limited. The 'genesis block', or the first block on the blockchain, is noteworthy. It has no hash that refers to a parent block, and it does not allow any mining process. Blocks are issued at fixed intervals. In current Ethereum blockchain new blocks can be released at every 15 seconds interval. The merkel tree has three type of nodes: (i) Leaf Nodes ($H_1$, $H_2$, $H_3$, ... $H_n$) (ii)Intermediate Nodes ($H_1 || H_2$,...$H_{n-1} || H_{n}$ ) and (iii) Root Nodes ($H_1 || H_2 || ... || H_{n-1} || H_{n}$). These hashes are also used as the node's reference key. The leaf node $(L_{i})$, intermediate node $(I_{i})$, and root node (R) of the MPT are defined as in equations \ref{eq2},\ref{eq3} and \ref{eq4} respectively.
\begin{equation}\label{eq2}
L_{i}= H_{i}=hash( x_{i} ), \big\{i \epsilon 1,2,3,4,....,m\big\} 
\end{equation}
\begin{equation}\label{eq3}
I_{i}= \big\{ H_{i} \parallel H_{i+1} \big\}   
\end{equation}
\begin{equation}\label{eq4}
 R= \big\{ H_{i} \parallel H_{i+1}... \parallel H_{N},N= Depth of MPT \big\}  
\end{equation}

\subsection{Delegated Proof of Stake Consensus Mechanism}
The Delegated Proof of Stake (DPoS) \cite{qin2018overview} consensus algorithm is a variance of the PoS mechanism which improves scalability and efficiency by lowering and limiting the number of validators on the network. It was designed to address the issue $scalability~trilemma$ \cite{qin2018overview}. In blockchain terminology, the more number of transactions per unit time refers to more scalability. As per the blockchain trilemma, more scalability may cause more challenges for security and decentralization features. In DPoS, token holders do not work on the validity of the blocks directly; instead, they choose delegates to validate transactions on their behalf. There are typically 21–100 designated delegates in a DPoS system. The chosen delegates are rotated regularly and the nodes order the delegates to present their blocks. When there are fewer delegates, it is easier to allocate one validator and time slot for each transaction. If the delegates consistently miss to validate transactions or blocks, it will cause erroneous transactions. As a result, the token holders vote them out and replace them with another delegate chosen by the token holders. 

\section {Working Principle}\label{work}
Once it receives the transaction request, the participant module starts its execution to check the status of the node. If the node is found as a new node, the required parameters such as the $PID$, $ETH~Address$, $keypair$, and a threshold value for $RID$ will be specified for the node. This $RID$ is incremented by one in each revocation call and once it reaches to the threshold the $PID$, $ETH~Address$ and $keypair$ of the node will be reset. The $PID$ and $ETH~Address$ identify a particular node at any time uniquely. \par
After the successful execution of the participant module, the smart contract is invoked. Thereafter the $PID$ of the selected signer node is compared with the stored $PID$. If both $PID$s are matched further execution will be allowed otherwise the process will be aborted. Then the $RID$ counter will be compared with its threshold limit. If the $RID$ exceeds the given threshold, the transaction will be aborted immediately otherwise, the signature module will be invoked. The adoption of the smart contract in our methodology helps the network to deal with the DDoS and MITM attacks by verifying the node id and checking the limit respectively.\par
The signature module allows the selected signer node to validate the transaction by verifying the public key. The Signature module allows that particular node to validate the transaction which completes the smart contract verification phase.\par
\begin{figure}[h!]
\centering
\includegraphics[scale=0.34]{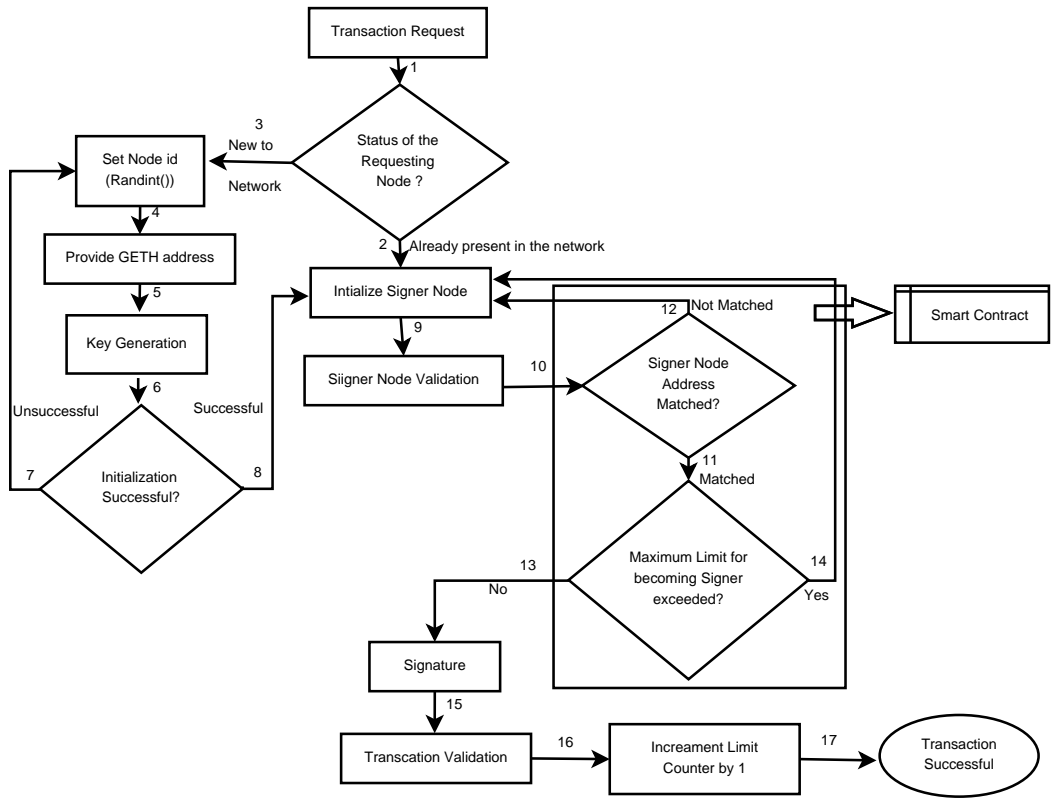}
\caption{Workflow of proposed Blockchain based PKI} 
\label{fig:flow}
\end{figure}

%Executing the Signature module enables the flagged signer node to validate the transaction by verifying the public key. The Signature module allows that particular node to validate the transaction which completes the smart contract verification phase.\par
After every successful transaction, the signature revocation module is invoked where the signer node increments its $RID$ value by 1 and validates the transaction. Figure \ref{fig:flow} represents the workflow of the proposed work.

\section {Implementation \& Comparison} \label{implement}
The proposed work can be evaluated by its performance and comparison with existing literature.
\subsection{Implementation}
The proposed work is implemented in the open-source Ethereum virtual machine  $GETH$. To invoke the smart contract, the $Solidity~v0.4.24$ scripting language is used along with the $GANACHE~truffle$ suit. The $truffle$ suit deploys the developed smart contract in the blockchain environment. Initially, the Gas limit of the network is set as $4000000$ and all created nodes have $100ETH$ in their account. The experiment is carried out with a Windows 10 OS, 8 GB RAM, 1 TB HDD, and $Intel~i5$ processor with a 2.8GHz clock speed machine.
\vspace{-1em}
\subsection{Performance}
\begin{figure}[h!]
\begin{minipage}{0.5\textwidth}
\centering
\includegraphics[scale=0.5]{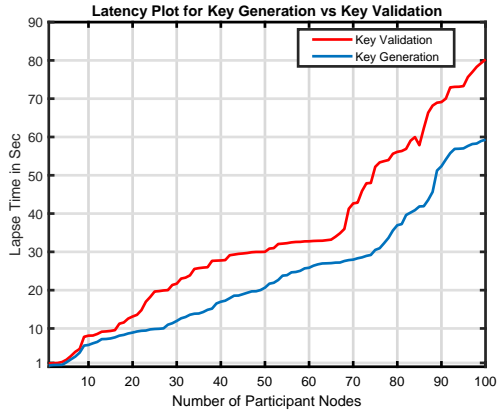}
\caption{Latency vs Number of Nodes for Key generation and Key validation Processes} \label{fig:latkey}
\end{minipage}
%\end{figure}
%\vspace{-1cm}
%Figure 2
%\begin{figure}[h!]
\hfill
\begin{minipage}{0.5\textwidth}
\centering
\includegraphics[scale=0.5]{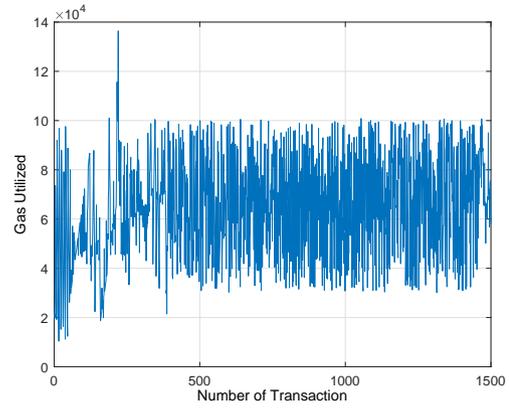}
\caption{Gas utilization vs Number of Different Transactions in The Network} \label{fig:gasnode}
\end{minipage}
\end{figure}

 The performance of the proposed PKI system is evaluated using the latency and gas utilization during the transaction. Figure \ref{fig:latkey} shows the latency vs the number of nodes graph for key generation and key validation process. The proposed model is tested with 100 nodes where latencies of key generation and key validation process reach to 60 seconds and 80 seconds respectively which is suitable for realistic applications of PKI. Figure \ref{fig:gasnode} shows gas utilization vs the number of transaction graph where the average gas cost for each transaction is approximately $10\times 10^4$. Table \ref{tab:gascost} shows the gas used by the different modules of the developed PKI system for doing one transaction.
%Table 1
\vspace{-1em}
\begin{table}[h!]
%\begin{center}
\centering

\caption{Gas usage by various modules } \label{tab:gascost}

\begin{tabular}{|p{3cm}|p{3cm}|p{3cm}|}

\hline

Method &  \multicolumn{2}{c|} {Gas Utilized} \\
\cline{2-3}

Name & For Initialization &  For Transaction \\
\hline
Participant & 33781 & 17484 \\
\hline
Signature &  42856 & 13752\\
\hline
Revocation &  19798 & 9689\\
\hline
Smart Contract  & 194837 & 32675 \\
\hline

\end{tabular}
%\end{center}

\end{table}
%Figure 1

\vspace{2em}
\subsection{Time Complexity Evaluation}
There are four executable modules present in the developed blockchain based PKI system, namely $participant$, $signature$, $revoke$, and $smart~contract$. Among these four modules, the time complexity of $participant$ and $smart~contract$ module is $O(n)$, whereas the time complexity of $signature$ and $revoke$ modules are $O(1)$. Here $n$ is the number of transactions committed to the procedure in the network. Multiple transaction requests may be raised in the case of participant and smart contract module resulting in the worst time complexity of these two modules as $O(n)$. While there is no communication in the other two modules: $signature$ and $revoke$ and also, no acknowledgment messages are issued to the transaction initiator.  The $signature$ and $revoke$ modules allow the chosen signer node (by smart contract module) to sign the transaction and make an increment of $RID$. So these two procedures do not generate any transaction messages, which results in constant time complexity of $O(1)$. Implementing the DPoS consensus mechanism results in a run time complexity of $O(logn)$. The time complexity of the whole system is $O(n+logn)$. The time complexity of the proposed model is compared with the different exiting models in Table \ref{tab:timecom}.

\begin{table}[h!]
\centering
\caption{Module wise Time Complexity Comparison with different existing models} \label{tab:timecom}

\begin{tabular}{|p{3cm}|p{3cm}|p{3cm}|p{3cm}|}

\hline
Blockchain based PKI &	Key/ Certificate Generation &	Key/ Certificate Validation &	Signature/ Certificate Revocation\\
\hline
PA-PKI \cite{wang2018privacy} &--&	$O (n)$	& $O (n)$ \\
\hline
CERT- CHAIN \cite{chen2018certchain} 	& $O (n^2)$&	--	&$O(log(n))$ \\
\hline
CERT-LEDGER  \cite{kubilay2019certledger} & $O (log(n))$ & -- &	--\\
\hline
DB-PKI \cite{toorani2021decentralized} &	$O (n^2)$ &	-- &$O (n^2)$\\
\hline
FLY-CLIENT \cite{bunz2020flyclient} &	--	 &$O (logn)$ &	--\\
\hline
BLOCKQUICK \cite{exosite2019blockquick} &	--	 &$O (n)$ &	--\\
\hline
Proposed System	& $O (n)$ &	$O (1)$	 & $O (1)$ \\
\hline
\end{tabular}

\end{table}
\vspace{-2em}
\subsection{Critical Analysis}
This work addresses various limitations of the existing PKI solutions including PKI without blockchain stated in Sec. \ref{without} and PKI with Blockchain stated in Sec. \ref{with}. The PKI provided in \cite{CT} only focuses on making the issued certificate visible to the network participants but does not have any circumstances to avoid the single point of failure (SPoF) limitation. In AKI \cite{kim2013accountable} the ILS is responsible to store the certificate issued by $CA$ and the ILS will be updated at a given time interval even $CA$ becomes untrusted. This becomes the key limitation along with SPoF as it is using a centralized $CA$ to issue the certificate. ARPKI \cite{basin2014arpki} tries to solve the synchronization issue of AKI but it still depends upon a trusted $CA$ to issue the certificate. The unavailability of the $CA$ verification process makes it tough to adopt ARPKI as a preferred solution.  The approach in PoliCert \cite{POLI} provides a centralized way to detect the log misbehavior which is again pruned to SPoF issue. LOCALPKI  \cite{dumas2017localpki} was created for usage in the context of IoT, where the local authority is in charge of utilizing the public key to verify the user's identity. The certificates issued by the local authority are stored by a third party, which are trimmed to SPoF. In NBPKI \cite{vigil2012notary} RA is in charge of authenticating the user's identification, and the NA maintains the user's status as trusted or untrusted based on the RA's decision. The malicious RA has the potential to compromise the system's integrity. \par
The $PA-PKI$ in article \cite{wang2018privacy} uses $Practical~Byzantine~Fault~Tolerance (pBFT)$ consensus model which allows a certain number of faulty nodes. If the number of faulty nodes exceeds that certain limits the whole network will be reset. Moreover, $pBFT$ consensus mechanism used in $PA-PKI$ of article \cite{wang2018privacy} and $DB-PKI$ of article \cite{toorani2021decentralized}  is prone to the $Sybil~attack$. The  $Block-CAM$ in article \cite{wang2018blockcam} has used the consortium blockchain platform for developing their PKI. The major limitation of using this platform is making the entire system semi-centralized since the consensus is managed by a certain number of participating nodes. Thus, it deviates from the decentralization concept of blockchain. Our proposed model is completely decentralized to all existing nodes in the network. The transactions of $BCTRUST$ in \cite{hammi2018bctrust}, depend upon the degree of trustworthiness of a participating node. Once the node is declared as the trusted one, then every node in the same network has to consider that node as the same. Moreover, all the transactions made by that node are also considered as valid transactions which may cause integrity loss and many other cyber threats. In our model verification is done on every transaction where node identity already padded, it does not verify only such node based identity. This feature makes our model more secure compared to \cite{hammi2018bctrust}. The implementation of PGP of both server and client-side participating nodes in $Block PGP$ \cite{yakubov2020blockpgp} causes heavy computational overhead which is the major drawback of such PGP based infrastructure. 

In the case of $PB~PKI$\cite{axon2016pb}, the transactions is stopped if any anonymous node requests to join the network. When new anonymous node requests to join the network, the entire network is disrupted until the joining request is processed. The developed PKI $TTA~SC$ in article \cite{ahmed2018turning} suffers from the loss of control issue over the blockchain network if it loses the key pair of the lead node under some cyber attacks. The DDoS attack to a particular lead node can make the system destabilized.  The $CERT~CHAIN$ in article \cite{chen2018certchain} uses $dependability~rank$ based consensus algorithm where the elected $CA$ was responsible for increasing the trustworthy degree. Depending upon the degree of trustworthiness the node will be elected as $CA$. Thus, a DoS attack on the particular $CA$ can cause damage in further transactions. It also uses the $ PBFT$ consensus algorithm which may cause a Sybil attack. In the $CERT~LEDGER$ of the article \cite{kubilay2019certledger}, the $CA$ is responsible for publishing the revoked data after every transaction. Thus, a DoS attack on that particular $CA$ can disrupt the entire network.
  The developed PKI $IKP$ in article \cite{matsumoto2016ikp} depends on the bitcoin’s language script which becomes hard to implement. The transaction process in $IKP$ depends on the trustworthiness of the $CA$ and if the $CA$ is misconfigured then all transactions within the network will be discarded. The $FLY~CLIENT$ of article \cite{bunz2020flyclient} does not use an authentication process to validate the participating node identification. So, the developed PKI is prone to $MITM$ attack.  In $BLOCK~QUICK$ of the article \cite{exosite2019blockquick}, the malicious block can only be detected by using the consensus group score. So, for a single malicious node, the whole branch will be discarded which will reduce the efficiency of the network.

\subsection{Attack and Defense}
The primary feature of the developed blockchain based PKI is the smart contract where the conditions such as the validity and threshold of the signer node are verified. The smart contract is solely responsible to allow the signer nodes to validate the key pair of requested nodes otherwise the nodes will be rejected. This feature avoids the DDoS and MITM attacks for the developed PKI. The proposed permissionless blockchain environment on the $GETH$ platform adopts the trust model of WoT where nodes are allowed to choose their own $CA$. 
In the hierarchical trust model, the processing power required to calculate the NONCE is high, whereas WoT does not require any NONCE calculation. The NONCE calculation can prevent MITM attacks. However, we have avoided it intentionally in our PKI to make it lighter compared to existing literature. From the storage point of view, only the hash value of each node is considered for the on-chain storage and the entire data is considered for the off-chain storage. Different attacks addressed in the current blockchain based PKI are reflected in Table \ref{tab:attack1}. Table \ref{tab:attack}reflects the various attack resistance comparison of the proposed model in contrast to other existing blockchain PKI models.

\begin{table}[h!]
\centering
\caption{Different Threats \& Its Defence} \label{tab:attack1}

\begin{tabular}{|p{1cm}|p{5cm}|p{6cm}|p{2cm}|}

\hline
Attack &	Basic Definition &	Prevention Mechanism &	Sustainability\\
\hline
DoS & The elected $CA$ may initiate a huge number of transactions. &The proposed model defines a threshold for every node for becoming a $CA$ and the implemented smart contract checks the given threshold with the $RID$. If the $RID$ exceeds the threshold, the participation of $CA$ will be rejected (see. \color{blue} Algo. \ref{algo:smart}).& Moderate \\
\hline
DDoS 	& Multiple elected $CA$s overload the network by initiating multiple transactions. &	The nodes in the network can become a $CA$ if the $RID$ is less compare to the given threshold. (\color{blue} Algo. \ref{algo:smart})& Moderate \\
\hline
MITM & An intermediate node may try to modify the transaction. This can be done in two ways such as modifying the content or modifying the sender/receiver node address. & Hash prevents content modification. The node verification process at the smart contact resolves the address violation part.  &low\\
\hline
51\%  &It is an attack on blockchain where attackers acquire the control of more than 50\% of the network's node address and cause faulty transactions. &	Before initiating the transaction, the node identity will be checked and only the active node of the network will be allowed for the transaction. (\color{blue}Algo. \ref{algo:part})  & low\\
\hline
Injection &	Injecting multiple unknown nodes to access the data &The adopted WoT does not allow the joining of a random node in the network.&low\\
\hline
Routing	& Tampering the data during the transaction &	Hashing is used to secure the information. & low \\
\hline
Eclipse attack	& The attacker may have a distributed botnets for replacing the actual node addresses by the false addresses.&For every transaction, the $PID$ will be checked for availability and WoT model restricts random joining of nodes & low \\
\hline
\end{tabular}

\end{table}
\begin{table}[h!]
\centering
\caption{Attack resistance comparison} \label{tab:attack}

\begin{tabular}{|p{3cm}|p{1cm}|p{1.1cm}|p{1cm}|p{1.2cm}|p{1.5cm}|p{1.5cm}|p{1.5cm}|}

\hline
PKI  						&DoS 	&DDoS &MITM&51\%&Injection&Routing&Eclipse \\
\hline
PA-PKI \cite{wang2018privacy} &	{\text{\ding{55}}}&{\text{\ding{55}}}&\checkmark&{\text{\ding{55}}}&{\text{\ding{55}}}&{\text{\ding{55}}}&{\text{\ding{55}}}\\
\hline
Block-CAM	\cite{wang2018blockcam}& \checkmark	&\checkmark	&\checkmark	&{\text{\ding{55}}}	&{\text{\ding{55}}}	&{\text{\ding{55}}}	& {\text{\ding{55}}}\\
\hline
BC-TRUST	\cite{hammi2018bctrust} & \checkmark	&\checkmark	&\checkmark	&{\text{\ding{55}}}	&{\text{\ding{55}}}	&{\text{\ding{55}}}	&{\text{\ding{55}}}\\
\hline
BLOCK-PGP \cite{yakubov2020blockpgp}&{\text{\ding{55}}}	&{\text{\ding{55}}}	&\checkmark	&\checkmark&{\text{\ding{55}}}	&{\text{\ding{55}}}	&{\text{\ding{55}}}\\
\hline
PB-PKI \cite{axon2016pb}&	\checkmark			&{\text{\ding{55}}}	&\checkmark	&{\text{\ding{55}}}	&{\text{\ding{55}}}	&{\text{\ding{55}}}	&{\text{\ding{55}}}\\
\hline
TTA-SC \cite{ahmed2018turning}&	{\text{\ding{55}}}	&{\text{\ding{55}}}	&\checkmark		&\checkmark	&{\text{\ding{55}}}	&{\text{\ding{55}}}	&{\text{\ding{55}}}\\
\hline
CERT-CHAIN \cite{chen2018certchain}&{\text{\ding{55}}}	&{\text{\ding{55}}}	&\checkmark	&{\text{\ding{55}}}	&{\text{\ding{55}}}	&{\text{\ding{55}}}	&{\text{\ding{55}}}\\
\hline
CERT-LEDGER \cite{kubilay2019certledger}&{\text{\ding{55}}}	&{\text{\ding{55}}}	&\checkmark	&{\text{\ding{55}}}	&{\text{\ding{55}}}	&{\text{\ding{55}}}	&{\text{\ding{55}}}\\
\hline
DB-PKI \cite{toorani2021decentralized} &{\text{\ding{55}}}	&{\text{\ding{55}}}	&\checkmark	&{\text{\ding{55}}}	&{\text{\ding{55}}}	&{\text{\ding{55}}}	&{\text{\ding{55}}}\\
\hline
IKP	\cite{matsumoto2016ikp}&{\text{\ding{55}}}		&{\text{\ding{55}}}	&\checkmark	&{\text{\ding{55}}}	&{\text{\ding{55}}}	&{\text{\ding{55}}}	&{\text{\ding{55}}}\\
\hline
FLY-CLIENT \cite{bunz2020flyclient} &	{\text{\ding{55}}}	&{\text{\ding{55}}}	&\checkmark	&{\text{\ding{55}}}	&{\text{\ding{55}}}	&{\text{\ding{55}}}	&{\text{\ding{55}}}\\
\hline
BLOCKQUICK \cite{exosite2019blockquick} &{\text{\ding{55}}}	&{\text{\ding{55}}}	&\checkmark	&{\text{\ding{55}}}	&{\text{\ding{55}}}	&{\text{\ding{55}}}	&\checkmark\\
\hline
Proposed System	& \checkmark				&\checkmark	&\checkmark	&\checkmark	&\checkmark	&\checkmark	&\checkmark\\
\hline
\end{tabular}

\end{table}

\section{Conclusion}\label{conclusion}
The proposed research work identifies several issues of conventional PKI and blockchain based PKI. In this regard, this work proposes a blockchain based PKI which is assisted by a smart contract and DPoS consensus algorithm. This work explores different existing solutions such as log based PKI, web of trust (WoT), and the blockchain based PKI system to deal with the various limitations and cyber threats of existing PKIs. The primary objective of this work is to create a blockchain based decentralized public key infrastructure which takes advantage of both the blockchain transparency and the web of trust model. The inclusion of smart contracts along with participant, signature and revoke modules in our work achieves the aforementioned features. The primary role of the adopted smart contract is used to validate the identity of the signer node and to check the threshold value for becoming $CA$. The DPoS consensus algorithm used in our PKI reduces the timing complexity of the transactions which makes our PKI affordable for lightweight applications. The performance of the proposed PKI system is evaluated based on the latency of the key generation, key validation, and signature revocation process. The gas utilization on the Ethereum platform is minimal for the initialization process and transactions. The proposed PKI can prevent DoS, DDoS, MITM, 51\%, Injection, Routing and Eclipse attacks. The developed smart contract used in our blockchain based PKI system is lighter to address the issue of storage limitation. 
%\nocite{*}% Show all bib entries - both cited and uncited; comment this line to view only cited bib entries;
\bibliography{wileyNJD-AMA}
\clearpage

\section*{Author Biography}

\begin{biography}{\includegraphics[scale=0.17]{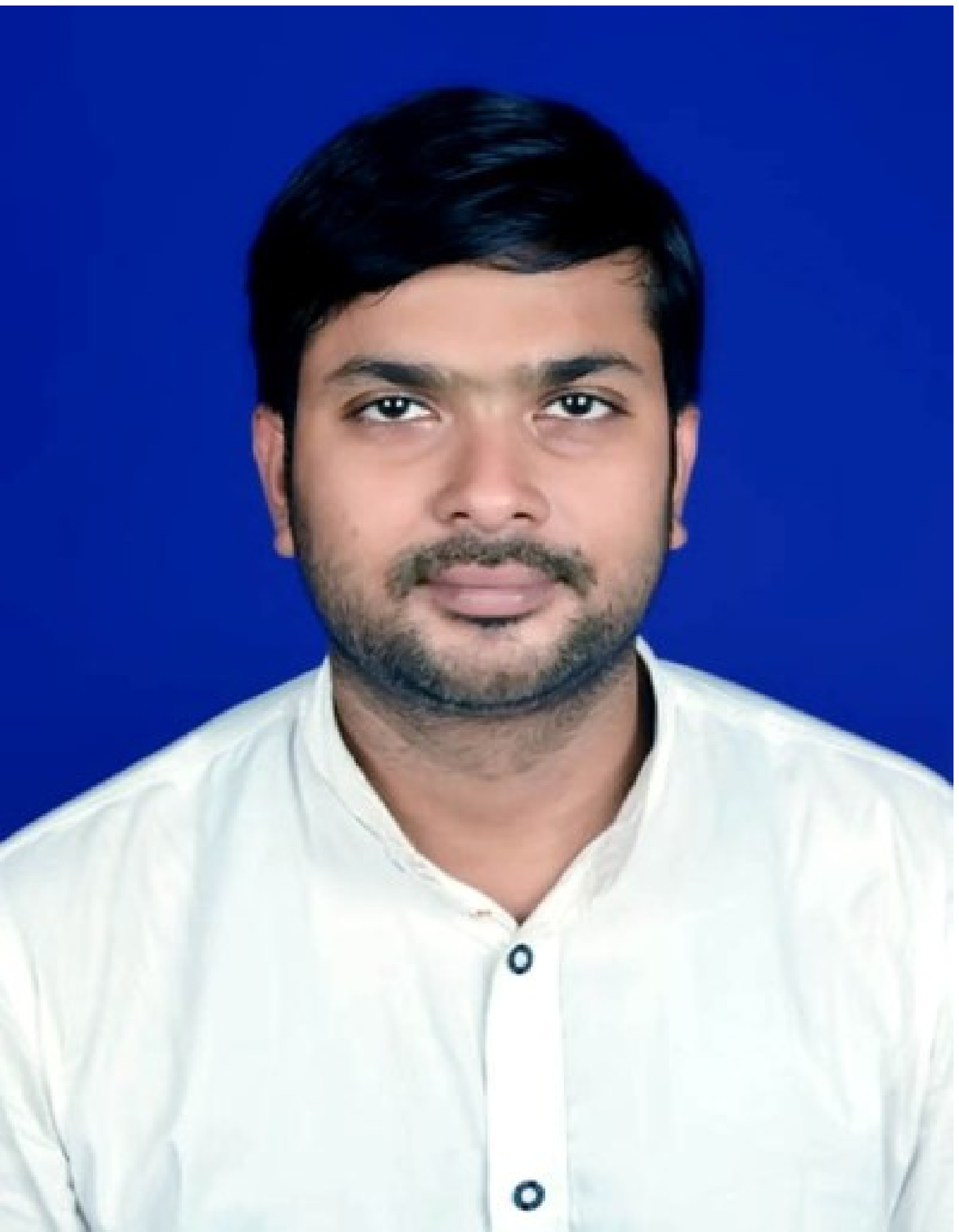}}{\textbf{Amrutanshu Panigrahi.} Mr. Amrutanshu Panigrahi is currently working as a research scholar in the department of CSE, Siksha O Anusandhan (Deemed to be University), Bhubaneswar, Odisha, India. He has obtained M.Tech in Information Technology from College of Engineering and Technology, Govt. of Odisha and B.Tech from BPUT Odisha. He is pursuing his PhD in the department of CSE at SOA University, Bhubaneswar. }
\end{biography}
\vspace{1cm}
\begin{biography}{\includegraphics[scale=0.17]{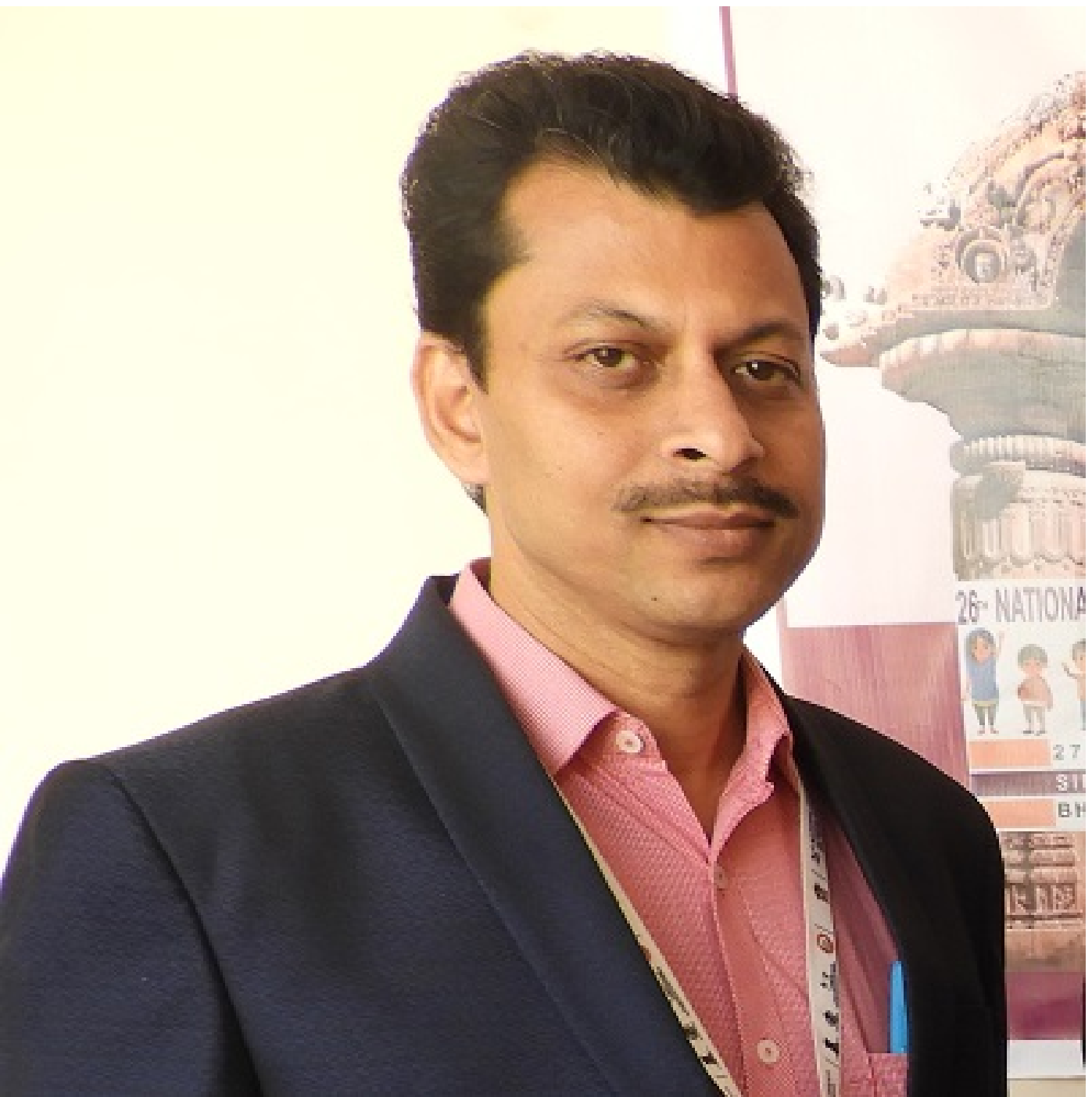}}{\textbf{Ajit Kumar Nayak.} Dr. Nayak is currently the professor and HoD of the Department of Computer Science and Information Technology, Siksha ‘O’ Anusandhan Deemed to be University, Bhubaneswar, Odisha. He graduated in Electrical Engineering from the Institution of Engineers, India in the year 1994, obtained M. Tech. and Ph. D. degree in Computer Science from Utkal University in 2001 and 2010 respectively. His research interests include Computer Networking, Ad Hoc \& Sensor Networks, Machine Learning, Natural Language Computing, Speech and Image Processing etc. He has published about 55 research papers in various journals and conferences. Also co-authored a book ‘Computer Network Simulation using NS2’, CRC Press. He has also participated as an organizing member of several conferences and workshops in International and National level.}
\end{biography}
\vspace{0.5cm}
\begin{biography}{\includegraphics[scale=0.18]{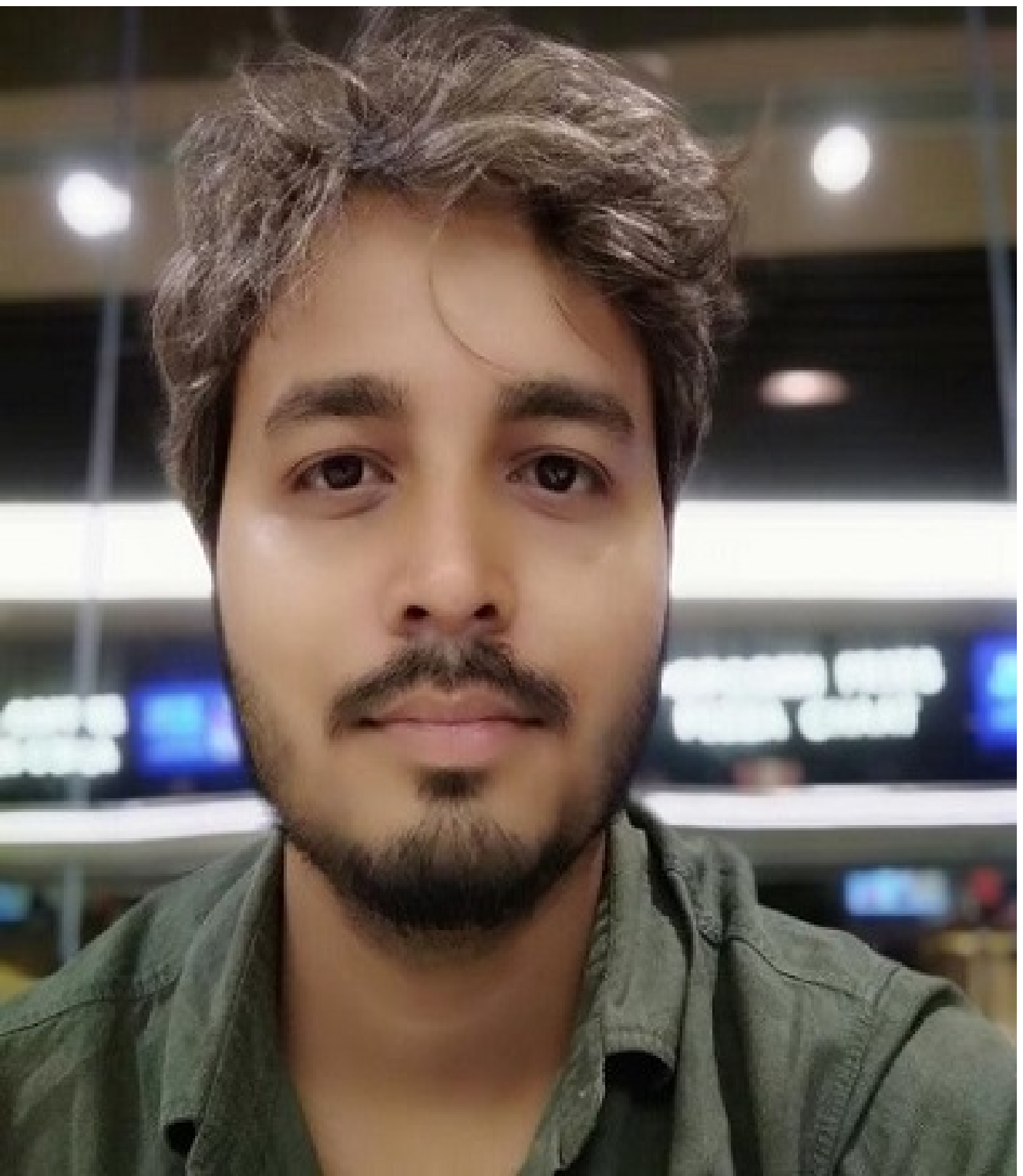}}{\textbf{Rourab Paul.} At present Dr. Paul is an Assistant Professor at Shiksha‘O’ Anusandhan University, Bhubaneswar, Odisha. Previously he was Post Doctoral Fellow in Computer Science and Engineering department of Indian Institute of Technology Kanpur. He has received his B.Sc.(2008) and M.Sc. (2010) degree with Electronic Science from University of Calcutta. He has completed his doctoral research on Cryptography and related VLSI design at the University of Calcutta, Kolkata, 2012–2017. He was a senior research fellow at the School of I.T. of Calcutta University. He also worked for the European Organization for Nuclear Research (CERN), Geneva, Switzerland in Large Ion Collider Experiment (ALICE) during 2015–2016. He held visiting research fellow position in Electronics and Communication Engineering department, National University of Singapore (NUS) during Sept 2013 to Jan 2014. He has been engaged in teaching, research and industrial consultancy from 2010 to 2012. He was visiting Lecturer in Acharya Prafulla Chandra College, Madhyamgram, Kolkata and Techno India, Saltlake, Kolkata, He was a senior academic consultant in Convergent Solutions, Saltlake, and also joined an internship program in i-cee Design Technology, Kolkata.}
\end{biography}

\end{document}